\documentstyle[preprint,aps]{revtex}
\tighten
\begin{document}
\draft
%__________________________________________________
%\twocolumn
%_______________________ Title, Authors ____________________________________
\preprint{\parbox[t]{85mm}{Preprint Numbers: \parbox[t]{48mm}
        {UNITU-THEP-15/1994\\ TU-GK-94-003}\\[2mm]}}
%         Bulletin Board: hep-ph/jjmmnnn\\}}
\title{Analytic Structure of the Quark Propagator in a Model with an
infrared vanishing Gluon Propagator}
\author{Axel Bender\footnotemark[1] and Reinhard
Alkofer\footnotemark[2]}
\address{\ \\[-1mm] \footnotemark[1]Physics Division, Argonne National
Laboratory,\\ Argonne, Illinois 60439-4843, USA\\[2mm]
\footnotemark[2]Institut f\"{u}r
Theoretische Physik, Universit\"{a}t T\"{u}bingen,\\
Auf der Morgenstelle 14, D-72076 T\"{u}bingen, FRG}
%\date{}
%
\maketitle
%-------------------------------------------------------------------
\begin{abstract}
The Dyson-Schwinger equation for the quark self energy is solved in rainbow
approximation using an infrared (IR) vanishing gluon propagator that
introduces an IR mass scale $b$. There exists a $b$ dependent critical
coupling indicating the spontaneous breakdown of chiral symmetry. If one
chooses realistic QCD coupling constants the strength and the scale of
spontaneous chiral symmetry breaking decouple from the IR scale for small $b$
while for large $b$ no dynamical chiral symmetry breaking occurs.  At
timelike momenta the quark propagator possesses a pole, at least for a large
range of the parameter $b$. Therefore it is suggestive that quarks are not
confined in this model for all values of $b$. Furthermore, we argue that the
quark propagator is analytic within the whole complex momentum plane except
on the timelike axis.  Hence the na\"{\i}ve Wick rotation is allowed.
\vfill
\begin{center}
{\bf to appear in Phys.~Rev.~D}
\end{center}
\vspace*{\fill}
\end{abstract}
\pacs{Pacs Numbers: 12.38.Aw, 11.30.Qc, 12.38.Lg}
%%%%%%%%%%%%%%%%%%%%%%%%%%%%%%%%%%%%%%%%%%%%%%%%%%%%%%%%%%%%%%%%%%%%%%%%%%%%%%
\section{Introduction}
New accelerators like CEBAF, MAMI-B and COSY will investigate hadron
observables at a scale intermediate to the low-energy region where hadron
phenomena mainly reflect the underlying principles of chiral symmetry and its
dynamical breakdown and to the high momentum region where the ``strong''
interaction has the appearance of being a perturbation on free-moving quarks
and gluons.  A reasonable theoretical description of intermediate energy
physics therefore has to satisfy at least the following requirements:
Formulated in terms of quarks and gluons it has to account for a mechanism of
dynamical chiral symmetry breaking (DCSB) and identify the related Goldstone
bosons in the measured spectrum of quark bound states. In addition, all the
correlation functions have to transform properly under the renormalization
group of Quantum Chromo Dynamics (QCD), {\it i.e.} at large momenta one
should recover the correct anomalous dimensions. Only if these requirements
are fulfilled the theoretical framework is fast in those grounds of the
theory of strong interaction that are well established. Furthermore, it is
desirable to formulate a microscopic picture of the cause of confinement.

In this context growing interest has been focused on the Euclidean
Dyson--Schwinger equations (DSE's) as a tool for developping and applying
non-perturbative methods in quantum field theories, a recent review on this
subject is given in Ref.\cite{rob94}.  Realistic models of QCD derived
through truncation of the DSE tower are {\em applicable in the whole
Euclidean momentum space} and, among others, have the following features:
Quark condensation gives rise to DCSB, the chiral Goldstone bosons are
identified with the lowest mass pseudoscalar quark-antiquark bound states,
perturbative QCD is matched at large momenta, and a study of the analytic
structure of the quark propagator provides some insight into the mechanism of
confinement.

The DSE's include the Bethe-Salpeter equations (BSE's) for bound states,
especially the ones describing physical mesons \cite{faddeev}. The BSE's are
coupled to the DSE's for the quark and gluon propagators, {\it i.e.}\ the
BSE's need as input at least the full renormalized quark and gluon
propagators which --- for calculating the meson mass spectrum, for instance
--- have to be known at complex Euclidean momenta far away from the real
axis. In principle, the quark and gluon propagators can be determined from
their DSE's. As these are coupled to the infinite tower of DSE's one is not
able to solve the equations exactly. Possible approximation schemes are based
on the truncation of the DSE tower: Some $n$-point functions like the gluon
propagator and the quark-gluon vertex are parametrized choosing {\it
Ans\"atze} such that certain requirements are fulfilled. Important
constraints are imposed by the discrete and continuous symmetries of QCD most
of which are formulated in Ward or Slavnov-Taylor identities. Additional
input may be taken from hadron phenomenology.

In this paper we investigate the analytic structure of the quark propagator
using the Euclidean DSE for the quark self energy in rainbow approximation
where the full quark-gluon vertex is replaced by the perturbative one. The
main ingredient for our study is a color diagonal gluon propagator
whose transverse part vanishes at zero four momentum:
\begin{eqnarray}
D^{\rm IR}_T (k^2) = \frac{k^2}{k^4+b^4} \label{stingl1}\; .
\end{eqnarray}
Herein $k$ denotes the (spacelike) gluon momentum and $b$ is an energy scale
to be determined. This form of the gluon propagator is motivated by very
distinct considerations.  One is based on the complete elimination of Gribov
copies within a lattice formulation of non-abelian gauge
theories\cite{zwanziger}: Landau gauge is fixed uniquely with the help of a
`thermodynamic parameter'. The resulting functional integral is dominated by
field configurations on the Gribov horizon.  This implies that the gluon
propagator vanishes for small momenta $k$ as $k^2$. In a couple of different
studies a gluon propagator of the form (\ref{stingl1}) is obtained by
considering a generalized perturbative expansion of the seven superficially
divergent proper vertices of QCD which are allowed to depend non-analytically
on the coupling constant \cite{haebel}. Furthermore, the Field Strength
Approach (FSA) to QCD supports the form (\ref{stingl1}) for being the
dominant part of the infrared quark-quark interaction\cite{lorenz}; however,
the color structure of the FSA gluon propagator is not diagonal. At
last, some recent lattice calculations obtained a gluon propagator which
allow a fit of the form (\ref{stingl1}) (but which do not rule out a fit
using a standard massive particle propagator)\cite{bernard}.

As the gluon propagator (\ref{stingl1}) has no poles for timelike momenta it
may be argued that it describes confined gluons. Instead, the gluon
propagator has poles on the imaginary axis, $k^2=\pm i b^2$, and therefore
the gluon `decays' after some time $\tau \approx 1/b$. This can be
interpreted as the gluons being screened. On the other hand, the gluon
propagator (\ref{stingl1}) is smaller than the perturbative one for all
momenta: ${k^2}/{(k^4+b^4)} < {1}/{k^2}$. This has drastic consequences for
DCSB as we will show: For the values of $b$ that are suggested by the lattice
studies of Ref.~\cite{bernard} chiral symmetry will be realized in the
Nambu-Goldstone mode for unrealistic large coupling constants only. We also
find that quarks are not confined in this model, at least for a large range
of model parameters $b$.

A similar study was undertaken by Hawes et al.\cite{hawes}. In this work the
gluon propagator (\ref{stingl1}) and a non-trivial quark-gluon vertex has
been used to calculate the quark propagator in the spacelike region.  Their
results are similar to ours, especially they find unconfined quarks. In
contrast to our work where the analytic continuation to timelike momenta is
performed explicitly and a pole signaling free massive quasi quarks as
asymptotic states is found, Hawes et al.\ use a test for confinement
employing a Fourier transformation of the quark Schwinger functions.  The
large time behavior of these Fourier transforms indicate asymptotic states of
massive deconfined quarks.

In Sec.~2 we present our model DSE for the quark self energy. We also discuss
the {\it Ansatz} Eq.~(\ref{stingl1}) for the gluon propagator in more
detail. In Sec.~3 we present the numerical solution for the quark self energy
at real Euclidean momenta and discuss the dynamical breaking of chiral
symmetry. In Sec.~4 we describe the analytic continuation of the quark self
energy with special emphasis on its pole structure and the Wick rotation from
Euclidean to Minkowski space. In Sec.~5 we conclude.

\section{Dyson-Schwinger Equation for the Quark Self Energy}
Our starting point is the DSE for the QCD quark two-point function
in Euclidean momentum space which can be written in terms of the quark self
energy $\Sigma (p)$. After renormalization it reads \cite{metric}:
\begin{eqnarray}
   \Sigma (p) = (1-Z_2)\gamma\cdot p + (Z_2 Z_m -1) m
                + Z_1\, C_F g^2\,\int
                      \frac{d^4q}{(2\pi)^4}\,
                      D_{\mu\nu} (p-q)\, \Gamma_\mu (p,q)\,
                      S(q)\, \gamma_\nu \label{sderen} \,
\end{eqnarray}
with $C_F = (N_C^2-1)/(2N_C) = 4/3$ for three colors and $g$ denoting the QCD
coupling constant. $m$ is the current quark mass.  $Z_1 (\mu^2, \Lambda^2)$,
$Z_2 (\mu^2,\Lambda^2)$ and $Z_m (\mu^2,\Lambda^2)$ are renormalization
constants depending on a regularization parameter $\Lambda$ and getting fixed
at the renormalization point $p^2 = \mu^2$.

The three Green's functions that appear on the r.h.s.\ of Eq.~(\ref{sderen})
are

\noindent
{\bf 1.} the quark propagator,
\begin{eqnarray}
S(q) = \frac{1}{\gamma\cdot q-m-\Sigma (q)}
     = \frac{1}{\gamma\cdot q\, A (q^2) - B (q^2)}
     = \frac{Z (q^2)}{\gamma\cdot q - M (q^2)}\, , \label{defquark}
\end{eqnarray}
with $B$ and $A$ denoting the scalar and vector parts of the quark self
energy $\Sigma$, respectively, and $Z$ and $M$ the field renormalization
function and the mass function, respectively;

\noindent
{\bf 2.} the gluon propagator
\begin{eqnarray}
D_{\mu\nu} (k) = \left( \delta_{\mu\nu} -
                        \frac{k_\mu k_\nu}{k^2}\right) D_T (k^2)
                        + \xi \frac{k_\mu k_\nu}{k^4}
                         \label{defgluon}
\end{eqnarray}
where only the transverse piece, $D_T$, is dressed due to vacuum polarization
and

\noindent
{\bf 3.} the quark-gluon vertex $\Gamma _\mu (p,q)$.

Of course, the gauge parameter $\xi$ appearing in
the gluon propagator~(\ref{defgluon}) is a function of the
renormalization point $\mu^2$, {\it i.e.}\ any redefinition of the
renormalization point leads to a new gauge. The only gauge which is not
affected by renormalization is the fix point $\xi = 0$, {\it i.e.}\ Landau
gauge \cite{MP78,PT84}.  In our studies we therefore concentrate on this
case.

\medskip

\noindent
For studying Eq.~(\ref{sderen}) one has to put in the full gluon propagator
and the full quark-gluon vertex; an exact investigation of the quark
DSE~(\ref{sderen}) therefore requires solving their DSE's, too.  To do so the
gluon 3-- and 4-point functions, the ghost self energy, the quark-quark
scattering kernel, and so on, have to be known.  Each of them again fulfills
its own DSE. We do not address the problem of solving all these DSE's
simultaneously but instead truncate the infinite tower of DSE's by making
{\it Ans\"atze} for both the gluon propagator as well as the quark-gluon
vertex.

Neglecting ghost contributions to the gluon vacuum polarization the transverse
part of the gluon propagator (\ref{defgluon}), $D_T (k^2)$, is closely
related to the running coupling constant of QCD, $\alpha (k^2)$:
\begin{eqnarray}
g^2\, D_T (k^2) = \frac{4\pi \alpha (k^2)}{k^2}
\, .\label{gluv}
\end{eqnarray}
In addition,
\begin{equation}
Z_1 (\Lambda_{\rm UV},\mu^2) = Z_2 (\Lambda_{\rm UV},\mu^2) \label{z1=z2}
\end{equation}
in Landau gauge; this is a result of the Ward identity that connects the
quark-gluon vertex to the inverse of the quark propagator.

Taking into account one-loop corrections to the QCD perturbation expansion
one possible smooth interpolation between IR and UV behavior is the
following {\it Ansatz} for the gluon propagator:
\begin{eqnarray}
g^2\, D_T (k^2) = D^{\rm IR}_T (k^2)\,
         \frac{\lambda}{\log \left(\tau+k^2/\Lambda^2_{\rm QCD}\right)}
          \label{uvgluon}
\end{eqnarray}
with $\Lambda _{\rm QCD}$ the QCD scale parameter and $\lambda$ the coupling
strength which is related to the anomalous dimension of the quark mass,
$d_M=12/(33-2N_f)$\cite{Pol76,Yn}: $\lambda = 4\pi^2 d_M$. However, in our
numerical studies we will treat $\lambda$ as an adjustable parameter.
Furthermore, we set $\tau \equiv e$ to ensure that the whole strength of the
infrared (IR) quark-quark interaction is carried by the function $D^{\rm
IR}_T (k^2)$. In the following we will call the second factor on the r.h.s.\
of Eq.~(\ref{uvgluon}) `ultraviolet (UV) improvement'. Note that in
Eq.~(\ref{uvgluon}) the four-momentum $k$ is assumed to be spacelike. We like
to mention that its continuation to large complex momenta is by no means
fixed by the behavior in the asymptotic spacelike region.  For example one
could replace
\begin{eqnarray}\label{replacement}
\frac{\pi d_M}{\ln \left( \tau + k^2/\Lambda_{\rm QCD}^2\right)}
\longrightarrow \frac{2\pi d_M}{\ln \left( \tau^2 + k^4/\Lambda_{\rm
QCD}^4\right)}
\end{eqnarray}
without altering the spacelike UV behavior. Indeed, both expressions are
quite similar even for small spacelike four-momenta.  On the other hand,
their analytic properties are very different.  While the function on the left
hand side has its cut on the timelike axis, \mbox{$k\in iI\!\! R$}, the one
on the right hand side has cuts for \mbox{$k\in \sqrt{\pm i}I\!\! R$}.

Of course, one can use a two- or a three-loop order running coupling of QCD
instead of the UV-improvement in Eq.~(\ref{uvgluon}). Nevertheless, even
these cannot provide information about the structure of the gluon propagator
at small spacelike momenta ($\leq$ 2~GeV). Hence, for proceeding a
parametrization of $D^{\rm IR}_T$ is unavoidable.

\medskip

\noindent
On the parametrization of the IR part of the gluon propagator a great deal of
work has been done during the last years. Several parametrizations have been
investigated and have supported the opinion that quark confinement is closely
related to infrared slavery or at least to a strong finite quark-quark
interaction in the infrared momentum regime. Nevertheless, as remarked in the
Introduction, there exist very distinct studies \cite{zwanziger},
\cite{haebel} (using Landau gauge) and \cite{lorenz} that propose
\begin{eqnarray}
D^{\rm IR}_T (k^2) = \frac{k^2}{k^4 + b^4}\label{stingl}
\end{eqnarray}
with a dynamical mass scale $b$ assumed to be proportional to $\Lambda_{\rm
QCD}$. In addition, Refs.~\cite{zwanziger,lorenz} even argue that $b\approx
\Lambda_{\rm QCD}$. This propagator has no K\"all\'en--Lehmann representation
\cite{BD65} indicating that the propagation of a free quasi gluon is not
possible. As the strength of the quark-quark interaction vanishes at small
momenta (corresponding to large distances in coordinate space) the solution
of the quark DSE ~(\ref{sderen}) with the gluon propagator (\ref{stingl}) as
input would exhibit a new mechanism for quark confinement {\it if any}.

\medskip

\noindent
For simplicity we use the rainbow approximation, {\it i.e.}\ we formally
replace
the dressed by the perturbative quark-gluon vertex in Eq.~(\ref{sderen}):
\begin{eqnarray}
\Gamma_\mu (p,q) = \gamma_\mu\, .\label{rainbow}
\end{eqnarray}
In this approximation the solutions of the Bethe-Salpeter equations
(BSE's) for color-singlet quark-antiquark bound states show the correct low
energy behavior as formulated in current algebra and partial conserved axial
current (PCAC) relations. For instance, the quark bound states carrying the
quantum numbers of the pseudoscalar mesons can be identified
with the Goldstone bosons related to the dynamical breakdown of the
$SU_R(N_f)\times SU_L(N_f)$ chiral symmetry of massless QCD
\cite{DS79}; in case of massive QCD the pion mass is determined by a
generalized Gell-Mann--Oakes--Renner relation. The rainbow and the separable
approximation \cite{Rob95} are the only known approximations that imply the
foregoing consistent formulation of quark--DSE and BSE for mesons.
Furthermore, the rainbow approximation (\ref{rainbow}) is equivalent to the
stationary phase approximation of the global color-symmetry model (GCM)
\cite{CR85}, an abelianized version of QCD with quark fields being
transformed to collective fields composed of quarks and anti-quarks and
carrying the quantum numbers of mesonic fields.

\medskip

\noindent
After inserting Eqs.~(\ref{stingl}) and (\ref{rainbow}) (together with or
without Eq.~(\ref{uvgluon})) into Eq.~(\ref{sderen}), using Eq.~(\ref{z1=z2})
and fixing the renormalization point, $\mu^2$, such that vector and scalar
part of the quark self energy obey the renormalization conditions
\begin{eqnarray}
A(p^2=\mu^2) & = & \alpha \; , \label{rena}\\
B(p^2=\mu^2) & = & \beta \label{renb}
\end{eqnarray}
with constants $\alpha$ and $\beta$ being fixed later, we get two integral
equations which couple the functions $A$ and $B$ that now depend on the
momentum
$s=p^2$, the renormalization point $\mu^2$, and the regularization
parameter $\Lambda^2_{\rm UV}$:
\begin{eqnarray}
A(s, \mu^2;\Lambda^2_{\rm UV}) & = & \alpha\, \frac{1-J_A [A,M]
     (s,\mu^2;\Lambda^2_{\rm UV})}{1-J_A [A,M] (\mu^2,\mu^2;\Lambda^2_{\rm
     UV})}\; ,
\label{a} \\
B(s,\mu^2;\Lambda^2_{\rm UV}) & = & \beta + \alpha\, \frac{J_B [A,M]
 (s,\mu^2;\Lambda^2_{\rm
     UV})-J_B [A,M] (\mu^2,\mu^2;\Lambda^2_{\rm UV})} {1-J_A
     (\mu^2,\mu^2,\Lambda^2_{\rm UV})}
\label{b}
\end{eqnarray}
with
\begin{eqnarray}
\lefteqn{J_A [A,M] (s,\mu^2;\Lambda^2_{\rm UV})} \nonumber\\ & = &
               \frac{\lambda}{6\pi^3}\int_0^{\Lambda^2_{\rm UV}}
               \frac{dr}{A(r,\mu^2;\Lambda^2_{\rm UV})[r +
                M^2(r,\mu^2;\Lambda^2_{\rm UV})]} \int_{-1}^1 dx I_A
(x;r,s)\;\Xi
               (x;r,s) \label{JA} \; ,
\\[.5ex]
\lefteqn{J_B [A,M] (s,\mu^2;\Lambda^2_{\rm UV})}\nonumber\\ & = &
               \frac{\lambda}{6\pi^3}\int_0^{\Lambda^2_{\rm UV}} \frac{dr
               M(r,\mu^2;\Lambda^2_{\rm UV})}{A(r,\mu^2;\Lambda^2_{\rm UV})[r
                + M^2(r,\mu^2;\Lambda^2_{\rm UV})]} \int_{-1}^1 dx I_B
                (x;r,s)\;\Xi (x;r,s)\label{JB}
\end{eqnarray}
where
\begin{eqnarray}
\Xi (x;r,s) = \left\{\begin{array}{l@{\hspace{.3cm}}l}
              1 & \mbox{without} \\
              \left[\log
                  \left(\tau+r+s-2x\sqrt{rs} /\Lambda^2_{\rm QCD}
                  \right)\right]^{-1}
                & \mbox{with UV improvement, Eq.~(\ref{uvgluon})} \, .
              \end{array}\right. \label{uvimprovement}
\end{eqnarray}
The functions $I_A$ and $I_B$ are defined in the appendix, Eqs.~(\ref{ia})
and (\ref{ib}), where also the analytic expressions for the angle integrals
($x$-integrals) in case of $\Xi = 1$ are given (Eqs.~(\ref{iaex}) and
(\ref{ibex})). The mass function in Eqs.~(\ref{a}) and (\ref{b}) reads:
\begin{eqnarray}
M (s,\mu^2;\Lambda^2_{\rm UV}) & = & \frac{B(s,\mu^2;\Lambda^2_{\rm
 UV})}{A(s,\mu^2;\Lambda^2_{\rm UV})} \nonumber \\ & = & \frac{\beta
 \left[{1-J_A (\mu^2,\mu^2,\Lambda^2_{\rm UV})}\right] +
 \alpha\left[J_B(s,\mu^2;\Lambda^2_{\rm UV}) - J_B(\mu^2,\mu^2;\Lambda^2_{\rm
 UV})\right]}{\alpha\left[ 1-J_A (s,\mu^2,\Lambda^2_{\rm UV})\right]}\;
 . \label{mass}
\end{eqnarray}

To make contact with perturbative QCD one has to formulate an appropiate
renormalization condition ensuring the asymptotic freedom of quarks. It is
convenient to choose the renormalization point as large as possible and set
\begin{eqnarray}
\label{renauv}\alpha & = & 1\; ,\\
\label{renbuv}\beta & = & m
\end{eqnarray}
at this point.  The largest available momentum is \mbox{$s = \Lambda^2_{\rm
UV}$}; therefore we set: \mbox{$\mu^2 \equiv \Lambda^2_{\rm UV}$} if not
stated otherwise.

Numerically, it turns out that $J_A [A,M] (\mu^2,\mu^2;\Lambda^2_{\rm UV})$
is very small for $\mu^2$ being large enough. Actually, in the limit
\mbox{$\mu^2\to\infty$} it vanishes exactly.  Hence, solving the
DSE~(\ref{a}) and (\ref{b}) for the renormalized quark self energy is (within
numerical accuracy) identical to solving the equations
\begin{eqnarray}
A (s, \mu^2;\Lambda^2_{\rm UV}) & = & 1 - J_{A}
[A,M] (s,\mu^2;\Lambda^2_{\rm UV})
\; ,\label{a0}\\
B(s,\mu^2;\Lambda^2_{\rm UV}) & = & \widehat{m}(\mu^2) + J_{B}
[A,M] (s,\mu^2;\Lambda^2_{\rm UV})
\; ,\label{b0}
\end{eqnarray}
when imposing the renormalization condition
\begin{eqnarray}
M (s,\mu^2;\Lambda^2_{\rm UV}) = \frac{m A(\mu^2, \mu^2;\Lambda^2_{\rm UV})
+ B (s,\mu^2;\Lambda^2_{\rm UV}) - B
(\mu^2,\mu^2;\Lambda^2_{\rm UV})}{A (s,\mu^2,\Lambda^2_{\rm
UV})}\; . \label{mass0}
\end{eqnarray}
$\widehat{m}$ is a running current mass
\begin{equation}
\widehat{m}(\mu^2) = m - J_{B} [A,M]
(\mu^2,\mu^2;\Lambda^2_{\rm UV}) \; .
\end{equation}

We like to remark that in case of the DSE system~(\ref{a0},\ref{b0})
this condition is sufficient because the approximation
\mbox{$J_A(\mu^2,\mu^2;\Lambda_{\rm UV})=0 $} sets the renormalization
constant \mbox{$Z_2 (\mu^2;\Lambda_{\rm UV}) = 1$}. For the
system~(\ref{a},\ref{b}) $Z_2$ has to be fixed dynamically and hence two
constraints, Eqs.~(\ref{rena}) and (\ref{renb}), are needed.

\medskip

\noindent
\underline{\bf Asymptotic behavior of the quark self energy}\hspace{4mm}

\noindent
At large spacelike momenta $s$ the IR mass scale $b$ in the gluon
propagator~(\ref{stingl}) can be neglected and the integral equations
(\ref{a}) and (\ref{b}) with \mbox{$\Xi =1$} ({\em i.e.\/} no UV improvement)
can be converted into differential equations \cite{fukuda}.  As for large $s$
the integral $J_A [A,M] (s,\mu^2,\Lambda_{\rm UV}^2)$ becomes very small the
vector part of the quark self energy is approximatly equal to 1 and hence
\mbox{$Z_2\approx 1$}. The scalar part $B$ is then determined by the
non-linear differential equation
\begin{eqnarray}\label{BDGL}
s\,\frac{d^2B}{ds^2} + 2\,\frac{dB}{ds} +
        \frac{\lambda}{4\pi^2}\frac{B}{s+B^2} = 0
\end{eqnarray}
with the boundary condition formulated in Eqs.~(\ref{renb},\ref{renbuv}).  If
there exists a bounded solution one can drop the term $B^2$ in the
denominator and one gets the asymptotic solution
\begin{eqnarray}
 \frac{B (s)}{B(0)} \;\;\; \stackrel{s\; \rm large}{-\!\!\! -\!\!\!
	-\!\!\!-\!\!\!-\!\!\!\longrightarrow}\;\;\; \frac{a}{\sqrt{s/\mu^2}}
	\left\{ \cos \left[ \frac{\kappa}{2} \ln\left( s/\mu^2 \right) +
	\varphi \right] \right\} \label{Mtail}
\end{eqnarray}
with $\kappa=\sqrt{\lambda /\lambda_C -1}$. $\lambda_C = \pi^2 \approx 9.87$
is (up to a trivial color factor \mbox{$C_F^{-1} = 3/4$}) the critical
coupling of Quantum Electro Dynamics (QED). Implying the UV boundary
condition~(\ref{renbuv}) one gets
\begin{equation}
\varphi = \arccos \frac{m}{a} \label{varphi}\, .
\end{equation}
The parameter $a$ is uniquely fixed by the IR boundary condition {\em i.e.\/}
by the solution of the integral equation~(\ref{b}) which determines the slope
of $B$ at the renormalization point $\mu^2$.

\medskip\noindent
In case of the UV improved model as defined by the gluon propagator
(\ref{uvgluon}) the asymptotic behavior of the mass function $M(s)$ differs
qualitatively from Eq.~(\ref{Mtail}). An analysis based on the
renormalization group yields
\cite{Pol76,rob94}:
\begin{eqnarray}\label{andim}
M(s) \;\;\; \stackrel{s {\rm\; large}}{-\!\!\! -\!\!\! -\!\!\!
                 -\!\!\!-\!\!\!\longrightarrow}\;\;\
       -\frac{\lambda}{3} \langle \bar{q} q\rangle_{\mu^2} \frac{1}{s}
        \frac{ \left[\log \left(\mu^2/\Lambda_{\rm
        QCD}^2\right)\right]^{-\lambda/4\pi^2}}{\left[\log\left(s/
        \Lambda_{\rm QCD}^2 \right) \right] ^{1-\lambda/4\pi^2}} \, .
\end{eqnarray}

\medskip

\noindent
{\tt \underline{\bf Dynamical Chiral Symmetry Breaking}\hspace{4mm}}

\noindent
Compared to the hadronic mass scale $\Lambda_{\rm QCD}$ the up and down quark
current masses $m$ are negligible.  In the limit of zero current masses,
$m=0$, the QCD Lagrangian is invariant under chiral transformations. The
Wigner-Weyl realization of the vacuum state corresponds to the trivial
solution of Eqs.~(\ref{a}) and (\ref{b}):
\begin{eqnarray}
B_{\rm trivial} (s,\mu^2,\Lambda_{\rm UV}^2)  = 0\; .\label{trivsol}
\end{eqnarray}
In QED in Landau gauge this trivial solution {\it maximizes} the CJT action
\cite{cjt} whose stationary phase condition is identical to the quark
DSE~(\ref{sderen}) \cite{craig86,atkinson87}.  This statement holds also true
for abelianized QCD as long as $A \approx 1$.  Hence, a vacuum configuration
with $B\not= 0$ will be dynamically favored if $A \approx 1$. This may be
tested by evaluating the CJT action at the stationary points.

A nontrivial solution of Eqs.~(\ref{a}) and (\ref{b}) (in the case $m=0$)
indicates the dynamical breaking of chiral symmetry, {\it i.e.}\ the
corresponding vacuum state is realized in the Nambu-Goldstone mode. From the
definition of the quark condensate
\begin{eqnarray}
 \langle \bar{q} q \rangle_{\mu^2} = -\lim_{x\to 0+} {\rm tr}
	[S(x,0)-S_0(x,0)] \stackrel{m=0}{=} -12 \int \frac{d^4p}{(2\pi)^4}
	\frac{B(p^2)}{A^2(p^2) p^2 + B^2(p^2)} \label{quarkcond}
\end{eqnarray}
it is obvious that a non-vanishing scalar part of the self energy, $B > 0$,
leads necessarily to a non-vanishing condensate $\langle \bar{q} q
\rangle_{\mu^2} < 0$. The scalar part of the quark self energy at zero
momentum, $B(0)$, or, correspondingly, the mass value $M(0)$ is therefore an
order parameter of DCSB.

\medskip

\noindent
\underline{\bf Confinement and the analytic continuation of the quark self
energy}

\noindent
The analytic continuation
of Eqs.~(\ref{a}) and (\ref{b}) to complex momenta is very important in
the studies of DSE's for the following reasons:

\noindent
{\bf 1.}
A quark propagator whose mass function obeys the relation
\begin{eqnarray}
s_M + M^2(s_M) = 0\label{mc}
\end{eqnarray}
and whose renormalization function~$Z$ is non-vanishing at $s=s_M<0$
describes a massive particle which should be detectable in some appropriate
experiment. Hence, one way to obtain information about quark confinement is
the examination of the pole structure of the quark propagator after
continuing real Euclidean to imaginary momenta, {\it i.e.} $s$ to $-s$: If
there does not exist any momentum $s_M$ for which Eq.~(\ref{mc}) is fulfilled,
there is no K\"all\'en-Lehmann representation for the quark propagator and
the described quasi-quark is confined.

\noindent
{\bf 2.} In order to solve the quark bound state equations ({\it i.e.}\ the
BSE's for mesons or the Fadde'ev equations for baryons) information about the
quark self energy in a large domain of the complex momentum plane is
needed. While external momenta ({\it e.g.}\ the momentum of a meson close to
its mass-shell) can be timelike the loop momentum occurring in the BSE is
spacelike. Hence, there appear combinations of spacelike and timelike momenta
as arguments of the quark self energy functions $A$ and $B$; the
analytic continuation of these functions to complex momenta is therefore
unavoidable if one wants to describe hadrons others than pions.

\noindent
{\bf 3.}  Studying the analytic properties of the quark self energy provides
some insight into the connection of the Euclidean formulation of QCD and QCD
in Minkowski space.  Recently it has been shown \cite{stainsby} that in
models containing both gluon and quark confinement the Wick rotation cannot be
performed na\"{\i}vely indicating thereby the close relation between
confinement and singularities or branch cuts appearing in the quark
propagator at some complex Euclidean momenta.

%%%%%%%%%%%%%%%%%%%%%%%%%%%%%%%%%%%%%%%%%%%%%%%%%%%%%%%%%%%%%%%%%%%%%%%%%%%%%%

\section{Quark Self Energy for Spacelike Momenta}
We have solved Eqs.~(\ref{a}) and (\ref{b}) by iteration with fixed cutoff
$\Lambda_{\rm UV}$. Different parametrizations and grid densities of the
discretized momentum interval $s\in [0,\Lambda^2_{\rm UV}]$ have been
used. All the results we report herein do not depend on changes of the grid
parametrization or further increasements of the grid density.

To check the accuracy of the numerical procedure we first study the
asymptotic behavior of the DSE-model without UV-improvement, {\it i.e.}\ we
set \mbox{$\Xi = 1$} in Eqs.~(\ref{a}) and (\ref{b}). At large
renormalization points we fix the mass function arbitrarily and fit a
tail of the form~(\ref{Mtail}) in the momentum region \mbox{$(b^2\ll
)\;\mu^2\le s\le \Lambda_{\rm UV}^2$} allowing all parameters $a$, $\varphi$
and $\lambda_C$ to get varied.

In Fig.~1 the function $B(s)/B(0)$ is shown for $\lambda=16$, $\mu=2^{14} b$,
$\Lambda_{\rm UV} = 2^{18} b$ and $m/B(0) = 0.0153$ together with an
asymptotic fit (\ref{Mtail}) with \mbox{$a = 0.06858$} and
\begin{eqnarray}
\varphi = 1.3461\;,\;\lambda_C = 9.869 \; .  \label{tailparas}
\end{eqnarray}
A comparison with the calculated values \mbox{$\lambda_C = \pi^2 \approx
9.870$} and \mbox{$\varphi = \arccos (m/a) = 1.3461$} shows high accuracy
for the numerical
solution of Eqs.~(\ref{a}) and (\ref{b}).

For \mbox{$\mu^2 \gg b^2$} (and \mbox{$\mu^2 \gg m^2$}) the value of the mass
function at zero momentum increases linearily with increasing renormalization
point,
\begin{eqnarray}
   	M(0) \propto \mu\; ,  \label{scaling}
\end{eqnarray}
a scaling behavior that is well known from studies of four-dimensional
QED \cite{fukuda}.  The vector part of the quark
self energy, $A$, tends toward unity while the scalar part $B$ admits
an almost constant value of order $\mu$ up to \mbox{$s\approx \mu^2 /
2$}. Obviously, the DSE-model defined by Eqs.~(\ref{stingl}) and
(\ref{rainbow}) is just an IR--regularised version of
four-dimensional QED for $b/\mu\ll 1$.

Modifying the asymptotic behavior of the gluon exchange by taking into
account the renormalization group (RG) improved loop-corrections, cf.\
Eq.~(\ref{uvgluon}), introduces a third mass scale: $\Lambda_{\rm QCD}$.
Hence, no scaling comparable to that of Eq.~(\ref{scaling}) occurs
anymore. Instead, the quark self energy functions $A$ and $B$ show the
proper RG transformation properties. For several parameter pairs
$(b/\Lambda_{\rm QCD},\lambda)$ we have studied the large momentum behavior
of the quark mass function $M$ and have found a very good agreement with
an asymptotic analysis based on the RG. This is illustrated in Fig.~2 where
the calculated mass function for \mbox{$\lambda = 30$}, \mbox{$b/\Lambda_{\rm
QCD} = 1$}, and \mbox{$\mu/\Lambda_{\rm QCD} = \Lambda_{\rm UV}/\Lambda_{\rm
QCD} = 2^{18}$} is shown and an asymptotic tail of the form
\begin{eqnarray}\label{Masy}
M(s) = \frac{c}{s} \left[\ln \left( \frac{s}{\mu^2}
\right)\right]^{d-1}
\end{eqnarray}
with
\begin{eqnarray}\label{Paraas}
c = 1.642\,\Lambda_{\rm QCD}^3\;\;\; \mbox{and}\;\;\; d=0.729
\end{eqnarray}
is fitted. The coupling constant $\lambda$ extracted from this UV tail
coincides within 5\% with the value inserted into the DSE
(\mbox{$d=\frac{\lambda}{4\pi^2}\approx 0.760$}) demonstrating good numerical
accuracy. (Note that $d$ is the exponent of a logarithmic term.)  The quark
condensate in the model characterized by the above parameter set can be
calculated from Eq.~(\ref{andim}) together with Eq.~({\ref{Paraas}}); its
value is
\begin{eqnarray}
\langle \bar q q\rangle _{\mu^2 = {\rm 1 GeV}^2} \approx
\left( -(0.70 \pm 0.03) \Lambda_{\rm QCD} \right) ^3
\approx -\left( (190 \pm 60)\, {\rm MeV}  \right) ^3\; .
\end{eqnarray}
We have used \mbox{$\Lambda^{(N_f = 4)}_{\rm QCD} = 270 \pm 80$}~MeV
\cite{particledata}. The same result is obtained calculating the trace of
the quark propagator, Eq.~(\ref{quarkcond}): \mbox{$\langle \bar q q\rangle
_{\mu^2 = {\rm 1 GeV}^2} \approx \left( -(0.67 \pm 0.02) \Lambda_{\rm QCD}
\right) ^3 $}.

In Fig.~3 the order parameter $M(0)$ is shown as a function of the coupling
constant $\lambda$.  For vanishing current mass, \mbox{$m=0$}, we have
studied the dependence on the renormalisation point~$\mu$ (here set equal to
the cutoff~$\Lambda_{\rm UV}$) choosing the IR mass scale $b$ equal to
$\Lambda_{\rm QCD}$ as suggested by Refs.~\cite{zwanziger,lorenz}.  The
critical behavior of the order parameter, \mbox{$M(0) \propto
(\lambda-\lambda_C)^{\beta}$} of the solution in the chiral limit gets
softened if the current quark mass is non-vanishing, {\em i.e.}\ we find the
well-known behavior of an order parameter of DCSB.  While increasing the
regularization parameter $\Lambda_{\rm UV}$ as well as the renormalization
point $\mu$ the critical coupling $\lambda_C$ decreases slightly.  The
extrapolation $\Lambda_{\rm UV}\to \infty$ yields a value of the critical
coupling of $\lambda_C = 28.98\pm 0.02$. As in case of $\lambda > \lambda_C$
the vector part of the quark self energy, $A$, is quite close to unity, the
Nambu-Goldstone phase of DCSB is dynamically favored.  The critical exponent
extracted from the data is $\beta = 0.59 \pm 0.01$. This value is almost
identical to $\beta = 0.57 \sim 0.58$ Hawes {\it et al.} have derived in
their study of the DSE system~(\ref{a0},\ref{b0}) inserting several {\it
dressed} quark-gluon vertices\cite{hawes}.

The critical coupling strongly depends on the IR scale~$b$: Small values of
$b/\Lambda_{\rm QCD}$ lead to small, large ones to large critical coupling
constants as one might expect.  For $N_F=0\;(2)$ the IR scale that yields a
critical coupling identical to the coupling suggested by the QCD asymptotics,
$\lambda_C\equiv \frac{48}{33} \pi^2$ ($\frac{48}{29} \pi^2$), is
$b_{N_F=0\;(2)}/\Lambda_{\rm QCD}\approx 0.26\pm 0.01\; (0.28\pm 0.02)$.
Therefore only for very small IR mass parameters $b \le 100$ MeV DCSB takes
place.

In Fig.~4 the mass functions for different IR parameters $b$ are shown for
$\lambda = \frac{48}{29} \pi^2$.  The parameter $b$ is varied by two orders
of magnitude (for $b/\Lambda_{\rm QCD} > 0.3$ no DCSB occurs).  It is
remarkable that neither the strength of DCSB as ``measured'' by the order
parameter $M(0)$ nor the momentum range that contributes dominantly to the
mechanism of DCSB depend on the IR scale $b$ introduced by the effective
gluon exchange, {\it i.e.}\ the scales of DCSB and the gluon mass scale $b$
are not or only loosely connected to each other if a realistic coupling
constant $\lambda$ is used.
%%%%%%%%%%%%%%%%%%%%%%%%%%%%%%%%%%%%%%%%%%%%%%%%%%%%%%%%%%%%%%%%%%%%%%%%%%%%%%

\section{Analytic Continuation of the Quark Self Energy}
\subsection{Mass Shell Condition and Constituent Quark Mass}\label{subsecmass}
To study quark confinement we will analyze the quark propagator for timelike
momenta. Therefore the Euclidean quark self energy has to be continued
from spacelike to timelike momenta~\mbox{$p \to i p$}. Correspondingly,
the variable $s=p^2$ has to be continued to negative values, $s<0$.

Eqs.~(\ref{a}) and (\ref{b}) can be understood as the defining equations for
the vector and scalar parts of the quark self energy, $A$ and $B$,
respectively. Therefore the analytic continuation can be performed by
smoothly changing the external momenta~$s$ while keeping the loop momenta~$r$
spacelike. Thus one generates the quark self energy as a function of complex
momenta by simply using complex $s$ in Eqs.~(\ref{a}) and (\ref{b}).
Obviously, $A^*(s^*)= A(s)$ and $B^*(s^*)=B(s)$ and hence solving the DSE in
the upper half of the complex plane is sufficient. Note that the functions
$A(r)$ and $B(r)$ inserted into the right hand side of Eqs.~(\ref{a}) and
(\ref{b}) are still those determined self-consistently on the spacelike
momentum axis.

First we generate the functions $A$ and $B$ on a small strip including the
positive real axis.  There we find that $A$ and $B$ change continuously. No
singularities appear and therefore the analytical continuation can be
performed patching open sets towards timelike momenta, \mbox{$s<0$}. Problems
arise caused by the poles of the angular integrals \mbox{$\int dx\, I_{A/B}
(x;r,s) \Xi (x;r,s)$}.  Those are kinematical singularities that appear for
\mbox{$s_0\geq\frac{b^2}{2 \sin^2 \vartheta /2}$} where \mbox{$s\equiv s_0
\exp (i\vartheta)$} with \mbox{$s_0 = |s|$}. They are integrable as can be
seen from the analytic expressions~(\ref{iaex}) and (\ref{ibex}) of the
appendix, for instance, or from a Laurent series expansion of the
kernels~(\ref{rena}) and (\ref{renb}).  Nevertheless, the numerical results
for the domain \mbox{$s_0\geq \frac{b^2}{2 \sin^2 \vartheta /2}$} have been
quite unstable against a change of numerical parameters.  Hence we restrict
our discussion on results for \mbox{$s_0 < \frac{b^2}{2 \sin^2 \vartheta
/2}$} only. Both functions, $A$ and $B$, are entire on this domain.

The identification of a mass pole in the quark propagator fullfilling
condition~(\ref{mc}) is only possible if \mbox{$b>\sqrt{2}
M(-{b^2}/{2})$}. This implies that in the case of DCSB our numerical
method of identifying such a pole is only reliable for IR mass scale
parameters $b$ larger than the created dynamical mass~$M(0)$. In this
restricted parameter space we find that the mass shell condition~(\ref{mc})
for the quark is fulfilled for both models, with and without UV--improvement.
{\em I.e.\/} close to the momentum $-s_M$ quarks propagate like stable
particles with constituent mass $M(-s_M)$. For parameter values \mbox{$b <
\sqrt{2} M(0)$} we are not able to make any statements about the existence of
a quark mass pole.

In Figs.~5a and 5b the mass function $M(s)$ in the UV-improved
model~(\ref{uvgluon}) for $(b/\Lambda_{\rm QCD},\lambda)=(0.27,\lambda_{\rm
QCD} = \frac{48}{29} \pi^2)$ and $(1,30)$, resp., are shown. The parameters
used are such that the kinematical singularities are far away from the
timelike momentum region considered here. One recognizes that the mass
condition~(\ref{mc}) is fullfilled and that the pole mass~$\sqrt{s_M}$ is
very well approximated by the dynamical mass~$M(0)$.

Our result of quark {\em deconfinement} (obtained for such values of the IR
parameter $b$ where our numerical results are stable)
is in full agreement with the results
of Hawes {\it et al.\/} They have investigated the scalar part of the quark
Schwinger function, $\Delta_S (T)$, analyzed its asymptotic behavior for
large Euclidean times~$T$, and extracted the mass $M_{\rm as}$ of a stable
asymptotic fermion state through the relation \cite{hawes}
$$\lim_{T\to \infty} \frac{d}{dT} \ln [\Delta_S (T) ] = -M_{\rm as}\, .$$

$M_{\rm as}$ is found to be (positive) finite and of the order of
$\Lambda_{\rm QCD}$. While we are using the rainbow
approximation~(\ref{rainbow}) for the quark-gluon vertex Hawes {\em et al.\/}
employ non-trivial vertices. These are constructed in a way that they obey
the Ward-Takahashi identity, carry no kinematical singularities, go versus
the free vertex for free fields, and ensure the propagators transforming
properly under the Landau-Khalatnikov transformations \cite{LK56,rob94}. In
addition, one of the vertices used in Ref.~\cite{hawes} leads to
multiplicative renormalizability \cite{CP90}.  Our study confirms the result
of unconfined quarks obtained for an IR vanishing quark--quark interaction
\cite{hawes}. We may also deduce that in this model the form of the
quark--gluon vertex is of minor importance (at least, as long as it is free
of singularities).

\subsection{Analyticity and Wick rotation}
In order to study the analytic properties of the quark propagator we employ
Cauchy's integral theorem in the following way. Suppose that the kernels
\begin{eqnarray}\label{fa}
f_A (r;s) &\equiv &\frac{A(r)}{rA^2(r) + B^2(r)}
               \int_{-1}^1 dx I_A (x;r,s)\;\Xi (x;r,s)\, , \\
f_B (r;s) &\equiv &\frac{B(r)}{rA^2(r)+ B^2(r)}
               \int_{-1}^1 dx I_B (x;r,s)\;\Xi (x;r,s)\, ,
\label{fb}\end{eqnarray}
of the integrals $J_A$ and $J_B$ (cf.\ Eqs.~(\ref{JA}) and (\ref{JB})) are
analytic. Then the integral of $f_{A(B)}$ over a closed contour in the
complex momentum plane has to vanish. Due to the renormalization
condition~(\ref{mass}) the functions $f_{A(B)}$ fall off faster than $1/r^x$
for large $r$ where
\mbox{$x > 1$}. Therefore one obtains
\begin{eqnarray}\label{ana}
\int_0^{\Lambda_{\rm UV}^2} dr f_{A(B)}(r) \approx \int^{\Lambda_{\rm
UV}^2}_0 dr\,e^{i\vartheta}
f_{A(B)}(r\,e^{i\vartheta})
\end{eqnarray}
where the difference between the left hand side and the right hand side
vanishes for \mbox{$\Lambda_{\rm UV}\to \infty $}.
This relation implies that the iteration of Eqs.~(\ref{a}) and (\ref{b})
along a rotated axis should lead to the same values $A(0)$ and $B(0)$ as the
iteration along the spacelike axis as long as the kernels~$f_{A(B)}$ are
analytic. This provides us with an analyticity test for the
functions~$f_{A(B)}$.

The gluon propagator~(\ref{stingl}) is symmetric under the exchange of
spacelike and timelike momenta:
\begin{eqnarray}\label{stinglsym}
D_T(-k^2) = -D_T(k^2)\, .
\end{eqnarray}
Even though the UV-improved version of this gluon exchange does not carry the
symmetry~(\ref{stinglsym}) an UV-improvement with the
replacement~(\ref{replacement}) does. As the spacelike properties (such as
dynamical chiral symmetry breaking) are
only slightly affected by the
replacement~(\ref{replacement}) we believe our symmetry considerations being
quite general and we concentrate on the UV-unimproved form~(\ref{stingl}) for
the gluon propagator alone.

In the following we study the approximate DSE system
(\ref{a0},\ref{b0}) and we drop the dependencies on regularization
parameter~$\Lambda_{\rm UV}$ and renormalization point~$\mu$.  Assuming $f_A$
and $f_B$ being analytic within the whole complex momentum plane the
iterations of Eqs.~(\ref{a0}) and (\ref{b0}) along the timelike axis
\begin{eqnarray}\label{zeita}
\widetilde{A} (-s) & = & 1 + \frac{\lambda}{6\pi^3}
\int_0^{\Lambda^2_{\rm UV}}
dr\,\frac{\widetilde{A}(-r)}{r\widetilde{A}^2(-r) - \widetilde{B}^2(-r)}
\int_{-1}^1 dx I_A (x;r,s) \, ,\\
\widetilde{B} (-s) & = & (m\to 0) + \frac{\lambda}{6\pi^3}
\int_0^{\Lambda^2_{\rm UV}}
dr\,\frac{\widetilde{B}(-r)}{r\widetilde{A}^2(-r) - \widetilde{B}^2(-r)}
\int_{-1}^1 dx I_B (x;r,s)
\label{zeitb}\end{eqnarray}
should lead to the same values~$A(0)$ and $B(0)$ as the iteration
along the spacelike axis.

Obviously,
\begin{equation}
\widetilde{A} (-s) = A(s)  \; ,\;\;\;\; \widetilde{B}(-s) = \pm i\,
B(s)
\label{zeitloes}
\end{equation}
are solutions of Eqs.~(\ref{zeita}) and (\ref{zeitb}) if $A$ and $B$
are solutions of Eqs.~(\ref{a0}) and (\ref{b0}). In the Nambu-Goldstone
realization of the vacuum, \mbox{$B(0)\not= 0$}, the
solutions~(\ref{zeitloes}) do not obey the analyticity condition
\mbox{$\widetilde{A} (0) = A(0)$} and \mbox{$\widetilde{B} (0) = B(0) $}.
Therefore the kernels $f_A$ and $f_B$ are {\em not\/} analytic in the whole
complex momentum plane in case of chiral symmetry being broken
dynamically. Note that a ``usual'' mass pole of the quark propagator is
sufficient to explain the behavior~(\ref{zeitloes}).

The functions $f_A$ and $f_B$ are products of the vector or scalar part of
the quark propagator and the angular integral over Euclidean projections of
the gluon propagator. We will argue in the following that the non-analyticity
of both functions are caused by the quark propagator. For doing this we study
the Wigner--Weyl phase of the system, \mbox{$B(s) \equiv 0$}. In that case
the self--consistency equation for \mbox{$A(s) = \alpha_1 (s_0,\vartheta) +
i\, \alpha_2( s_0, \vartheta )$} reads:
\begin{eqnarray}\label{atriv}
\alpha_1(s_0, \vartheta) + i\, \alpha_2 (s_0, \vartheta)  = 1 +
\frac{\lambda}{6\pi^2} \int^{\Lambda_{\rm UV}^2} dr\frac{1}{r(\alpha_1
(s_0, \vartheta) + i\, \alpha_2 (s_0, \vartheta))} \int_{-1}^1 dx\, I_A
\left(x;r,s_0 e^{i\,\vartheta}\right)
\end{eqnarray}
with $\alpha_1$ and $\alpha_2$ being the real and the imaginary part
of \mbox{$A\left(s=s_0\, e^{i\, \vartheta}\right)$}, respectively. It is
straightforward to proof the following symmetries
\begin{eqnarray}
\label{sym1}\alpha_{1/2}(s,\pi +\vartheta) & = & \alpha_{1/2} (s,\vartheta)\;
,\\
\alpha_1(s, \frac \pi 2 -\vartheta) & = & \alpha_1 (s,\vartheta)\; ,\\
\label{sym3}\alpha_2(s,\frac \pi 2 -\vartheta) & = & -\alpha_2 (s,\vartheta)\;
{}.
\end{eqnarray}
Hence it is sufficient to analyze the integral equation~(\ref{atriv}) for
\mbox{$\vartheta \in [0,\frac \pi 2 ]$} only.

In Fig.~6 the function $\alpha_1(s_0,\vartheta)$ is shown for \mbox{$\lambda =
16$} and various angles~$\vartheta$. We have used a small cut-off,
\mbox{$\Lambda_{\rm UV}^2 = 2^8 b^2$}. However, the solutions do not
depend on a further increasement of $\Lambda_{\rm UV}^2$ as the
chiral symmetric solutions do not display the scaling behavior~(\ref{scaling})
as in the case of the Nambu-Goldstone phase. The numerical error in Fig.~6 is
about 1\% to 5\%; it is largest close to the imaginary axis.

It is obvious that for all the angles~$\vartheta$ the real
parts~$\alpha_1$ at zero momentum (\mbox{$s =0$}) are identical; the
imaginary parts~\mbox{$\alpha_2$} which are not shown vanish for small
momenta within the numerical accuracy. For \mbox{$\vartheta = \frac \pi 2 =
90^o$}
the iteration of Eq.~(\ref{atriv}) does not converge due to the
(integrable) singularities of the gluon propagator. Nevertheless
there is no signature of any further non-analyticity close to the imaginary
axis. We therefore
conclude that there is a strong indication for the vector part of the
quark self energy being analytic. Then the quark propagator has only one
pole, namely the one at \mbox{$s=0$}.

In the Nambu-Goldstone phase, \mbox{$B(s) \not= 0$}, the integral kernels
$f_A$ and $f_B$ from Eqs.~(\ref{fa}) and (\ref{fb}) are not holomorphic on
the whole complex momentum plane either. There has to be a pole contribution
from the contour integration in Eq.~(\ref{ana}). As the angular integrals
over $I_A$ and $I_B$ do not carry a pole singularity the scalar and the
vector parts of the quark propagator have at least one pole. If this
singularity is not lying on the real axis but within the complex plane the
Wick rotation cannot be performed na\"{\i}vely. If it is placed on the
timelike axis the na\"{\i}ve Wick rotation is allowed as long as the causal
behavior of the quark propagator is suitably determined.

We have not found any singularities except the kinematical ones and the mass
pole on the real axis.  Even though we are not able to draw compulsory
conclusions this numerical result strongly indicates that the quark
propagator is analytic within the whole complex momentum plane except on the
real timelike axis.  This statement is also supported by the analysis within
different DSE models \cite{stainsby} where it is found that poles of the
quark propagators are lying quite close to the real axis.

In addition, the analyticity condition~(\ref{zeitloes}) is quite easily
understood from a physical point of view.  Under the action of the Wick
rotation the Euclidean mass term becomes \mbox{$(-i)$} times the
corresponding Minkowski term, a kinetic fermionic term keeps
``unmodified''. In terms of the quark propagator functions this reads:
\begin{eqnarray}
B(s_E) & \stackrel{\rm Wick}{\longrightarrow} & -i\,B(s_M)\label{wicka}\, ,\\
A(s_E) & \stackrel{\rm Wick}{\longrightarrow} &
A(s_M)\label{wickb}
\end{eqnarray}
where \mbox{$s_E$} denotes the Euclidean and \mbox{$s_M$} the corresponding
Minkowski four momentum squared close to a mass pole of the quark
propagator. In Minkowski space we therefore just recover one of the
``solutions''~(\ref{zeitloes}) which is not surprising.

Summarizing this section, there are strong indications that the Euclidean
quark propagator of the DSE model specified by the gluon
exchange~(\ref{stingl}) can be continued back to Minkowski space.  Within a
small region around the mass pole~(\ref{mc}) the Minkowskian quark propagator
has a K\"all\'en-Lehmann representation and it therefore resembles the
propagator of a massive stable fermion.  For a restricted set of parameters
we have been able to verify this numerically. Due to the numerical problems
associated with the kinematical singularities we are not really able to prove
this statement rigorously for all values of the parameter $b$. Nevertheless,
it is the most plausible conclusion drawn from the properties of the
functions~$f_A$ and $f_B$ as discussed in this section.

%%%%%%%%%%%%%%%%%%%%%%%%%%%%%%%%%%%%%%%%%%%%%%%%%%%%%%%%%%%%%%%%%%%%%%%%%%%%%%

\section{Summary and Conclusions}
We have investigated the Euclidean Dyson-Schwinger equation (DSE) for the
quark self energy for an infrared vanishing gluon propagator in rainbow
approximation. We have found that this model is not appropriate for
describing hadron phenomenology due to several reasons.  First, dynamical
chiral symmetry breaking (DCSB) occurs either for unphysically large coupling
constants only or for a small IR scale, $b\ll \Lambda_{\rm QCD}$. The
strength and the scale of DCSB are then independent of $b$: the spontaneous
breakdown of chiral symmetry is mainly driven by the term added to obtain the
correct UV behavior and {\em not} by the infrared part.

In addition, continuation to timelike momenta reveals that the quark
propagator possesses a pole. As our numerical method is only reliable for IR
mass scale parameters $b$ larger than the created dynamical quark mass (see
the discussion in subsection~\ref{subsecmass}) this pole can be unambigously
identified for such values of $b$ only. To be explicit, we do not claim that
there is no pole, we only state that within our method it cannot be found
even if it exists.

However, the occurrence of a mass pole in the propagator can be related to
the propagation of a stable particle. Therefore it is clear that quarks are
not confined in this model, at least for a large range of the IR
parameter~$b$.  This is in contradiction to results obtained in
Ref.~\cite{becker}. There heavy quarkonia have been studied using a
non-relativistic reduction of the gluon propagator~(\ref{stingl}) together
with a quark propagator and a quark-gluon vertex which result from the DSE
studies of Ref.~\cite{haebel}.  However, in this work the quark-quark
interaction~(\ref{stingl}) has been replaced by a Coulomb potential. The
resulting equation for the quarkonium bound states is just a
Lippmann-Schwinger equation for hydrogen with the exception that the
non--trivial quark propagators of Ref.~\cite{haebel} have been used.  These
propagators are then probably the reason for the ``confining mechanism''
described in Ref.~\cite{becker}. Stated otherwise, the unusual behavior of
the quarks have been put in ``by hand''.  Such a mechanism is not supported
by our calculations.

Instead, our results are in accordance with those of Hawes
et al.\ \cite{hawes} who have studied the quark propagator for spacelike
momenta
only and have extracted the quark mass from the large time behavior
of the Schwinger functions. Additionally, they have used
non-trivial quark-gluon vertices.  The good
quantitative agreement of our results with those of
Ref.~\cite{hawes} shows that the rainbow approximation is a good
approximation for this model interaction. This can be understood from
the fact that the influence of a non-trivial
quark-gluon vertex is concentrated in the infrared
which for the studied case is suppressed by the infrared vanishing
gluon exchange.

Even though a study of the analyticity of the quark propagator is hampered
by numerical difficulties
we have found convincing evidence for the quark propagator
being analytic within the whole complex Euclidean momentum plane
except the timelike (negative real) axis. This has two consequences:
First, it is possible to perform the na\"{\i}ve Wick rotation. Second, and more
important, the model does {\em not} lead to quark confinement.

\bigskip
%%%%%%%%%%%%%%%%%%%%%%%%%%%%%%%%%%%%%%%%%%%%%%%%%%%%%%%%%%%%%%%%%%%%%%%%%%%%%%

\acknowledgements
We take this opportunity to express our gratitude to
Prof.~Dr.~H.~Reinhardt for his support and his interest in
this work. We thank Dr.~C.~D.~Roberts for
numerous helpful discussions. AB got benefit
from the courses and meetings of the Graduiertenkolleg ``Hadrons and Nuclei''
(DFG contract Mu 705/3); therefore he especially
thanks Prof.~H.~M\"uther for organising these meetings.
He also gratefully acknowledges financial
support by the Studienstiftung des deutschen Volkes.
This work has also been partly supported by COSY under contract
41170833.
%\newpage

\bigskip
%%%%%%%%%%%%%%%%%%%%%%%%%%%%%%%%%%%%%%%%%%%%%%%%%%%%%%%%%%%%%%%%%%%%%%%%%%%%%%
\appendix
\section*{}
The angular integrands in Eqs.~(\ref{a}) and (\ref{b}) are:
\begin{eqnarray}
I_A (x;r,s) & \equiv & \sqrt{\frac{r^3}{s}} \sqrt{1-x^2}\,
               \frac{2(1+2x^2)\sqrt{rs}-3x(r+s)}{\left(r+s-2x\sqrt{rs}
               \right)^2 + b^4} \label{ia}\; , \\
I_B (x;r,s) & \equiv & r \sqrt{1-x^2}\,
               \frac{3(r+s-2x\sqrt{rs})}{\left(r+s-2x\sqrt{rs}
               \right)^2 + b^4} \; . \label{ib}
\end{eqnarray}
For $\Xi(x;r,s) = 1$ in Eqs.~(\ref{a}) and (\ref{b}) ({\it i.e.}\ no UV
improvement
for the quark-quark interaction) the x-integral can be evaluated exactly:
\begin{eqnarray}
\lefteqn{\int_{-1}^1 dx I_A (x;r,s) = } \nonumber \\
 & &\frac{\pi b^4}{4s^2} - \frac{\pi r b^2}{s y}
	\left\{b^2\left(\frac{(r-s)^2+b^4}{(r+s)^2+b^4}-\frac{(r+s)^2-b^4}{2rs}
     +  \frac{(r+s)^2((r-s)^2+b^4)}{rs((r+s)^2+b^4)}\right) w_+ \right.
        \nonumber \\
 & & \mbox{\hspace*{3.3cm}} +  \left. \frac{(r+s)((r-s)^2+5b^4)}{(r+s)^2+b^4}
	w_- \right\} \label{iaex}\, ,\\
\lefteqn{\int_{-1}^1 dx I_B (x;r,s) = \frac{3\pi (r+s)}{4s} } \label{ibex}\\
 & & - \frac{3 \sqrt{2} \pi r}{((r+s)^2+b^4) y} \left\{\frac{r+s}{4rs}
        \left((r-s)^4 +2b^2 (r+s)^2 +b^8 \right) w_+
  +b^2 ((r-s)^2-b^4) w_- \right\} \nonumber
\end{eqnarray}
with
\begin{eqnarray}
 y & = & ((r-s)^2+b^4)^2+16rsb^4+((r-s)^2+b^4)\sqrt{((r-s)^2+b^4)^2+16rsb^4}
	\, ,\\
 w_\pm & = & \sqrt{((r+s)^2+b^4)\sqrt{((r-s)^2+b^4)^2+16rsb^4}\pm
	(((r+s)^2+b^4)^2-4rs((r+s)^2-b^4) )}.\nonumber \\
\end{eqnarray}
%%%%%%%%%%%%%%%%%%%%%%%%%%%%%%%%%%%%%%%%%%%%%%%%%%%%%%%%%%%%%%%%%%%%%%%%%%%%%%

%%%%%%%%%%%%%%%%%%%%%%%%%%%%%%%%%%%%%%%%%%%%%%%%%%%%%%%%%%%%%%%%%%%%%%%%%%%%%%

%%------------------- Figures ---------------------------------
%%
\begin{figure}
\caption{The scalar part of the quark self energy,
$B(s)/B(0)$, is shown as a function of \mbox{$s=p^2$}.
The UV-unimproved gluon propagator~(\protect\ref{stingl}), the
coupling strength \mbox{$\lambda = 16$}, and the numerical cut-off
\mbox{$\Lambda_{\rm UV}^2 = 2^{36}b^2$} have been used. The solid line
represents the function obtained by iterating Eqs.~(\protect\ref{a}) and
(\protect\ref{b}). The dashed one is the asymptotic
tail~(\protect\ref{Mtail}) fitted at \mbox{$s=2^{28} b^2$} with the
parameters~(\protect\ref{tailparas}) inserted.}
\end{figure}
\begin{figure}
\caption{This figure shows the quark mass function $M(s)$ of the DSE-model
with UV-improved gluon propagator~(\protect\ref{uvgluon}). The parameters
used are: \mbox{$\lambda = 30$} and \mbox{$b/\Lambda_{\rm QCD} = 1$}.  The
solid line represents the function obtained by iterating
Eqs.~(\protect\ref{a}) and (\protect\ref{b}). The dashed one is the
asymptotic tail~(\protect\ref{Masy}) fitted at large momenta~$s$ with the
parameters~(\protect\ref{Paraas}) inserted.}
\end{figure}
\begin{figure}
\caption{In this figure the order parameter~$M(0)$ is shown as a
function of the coupling strength~$\lambda$ for the model with the
UV-improved gluon propagator~(\protect\ref{uvgluon}) and
\mbox{$b/\Lambda_{\rm QCD}= 1$}. The thick lines represent solutions of the
DSE~(\protect\ref{a}) and (\protect\ref{b}) for vanishing quark current mass,
\mbox{$m = 0$}, and different renormalization points~$\mu^2 = \Lambda_{\rm
UV}$. The thin lines are solutions for different current masses and fixed
\mbox{$\mu^2/\Lambda_{\rm QCD}^2 = \Lambda_{\rm UV}^2/\Lambda_{\rm QCD}^2 =
2^{36}$}.}
\end{figure}
\begin{figure}
\caption{In this figure the quark mass functions for different
IR parameters~$b$ are shown. The UV-improved gluon propagator is
used and the coupling strength is \mbox{$\lambda = \frac{48}{29} \pi^2$}.
The momentum range that contributes dominantly to the breaking of
chiral symmetry is defined as the value of $s$ where \mbox{$M(s)
=\frac{1}{2} M(0)$}.}
\end{figure}
\begin{figure}
\caption{The quark mass function at spacelike and timelike momenta for
\mbox{$(b/\Lambda_{\rm QCD},\lambda) = (0.27,\lambda_{\rm QCD} = {48}/{(29
\pi^2)})$} and $(1,30)$ are shown in Figs.~a and b, resp. The thin line on
the timelike side of the graphs is the function~$\protect\sqrt{-s}$.  The
point of intersection of this function with the mass functions gives the
value of $s$ where the pole condition~(\protect\ref{mc}) is fulfilled.}
\end{figure}
\begin{figure}
\caption{In this figure the real part~$\alpha_1$ of the scalar
quark self energy~$A$ is shown for the Wigner-Weyl phase of the DSE
model defined by the gluon exchange~(\protect\ref{stingl}). The
different functions correspond to solutions of Eq.~(\protect\ref{atriv})
iterated along different axes in the Euclidean momentum plane
characterized by the angle~$\vartheta$.}
\end{figure}

\begin{references}
\bibitem{faddeev}
Baryons are described by relativistic Fadde'ev equations.
\bibitem{metric} We use negative semidefinite Euclidean
metric, $g_{\mu\nu} = -\delta_{\mu\nu}$, $a\cdot b = -\sum_{j=1}^4 a_j
b_j \le 0$ $\forall a=b \in I\!\! R ^4$. For abbreviation, we
nevertheless define the square of a momentum as \mbox{$q^2 := -q\cdot q
\ge 0$}. All Euclidean Dirac matrices,
$\gamma_\mu$, are antihermitean and obey the anticommutation relations
$\{\gamma_\mu ,\gamma_\nu\} = -2\delta_{\mu\nu}$.
\end{references}
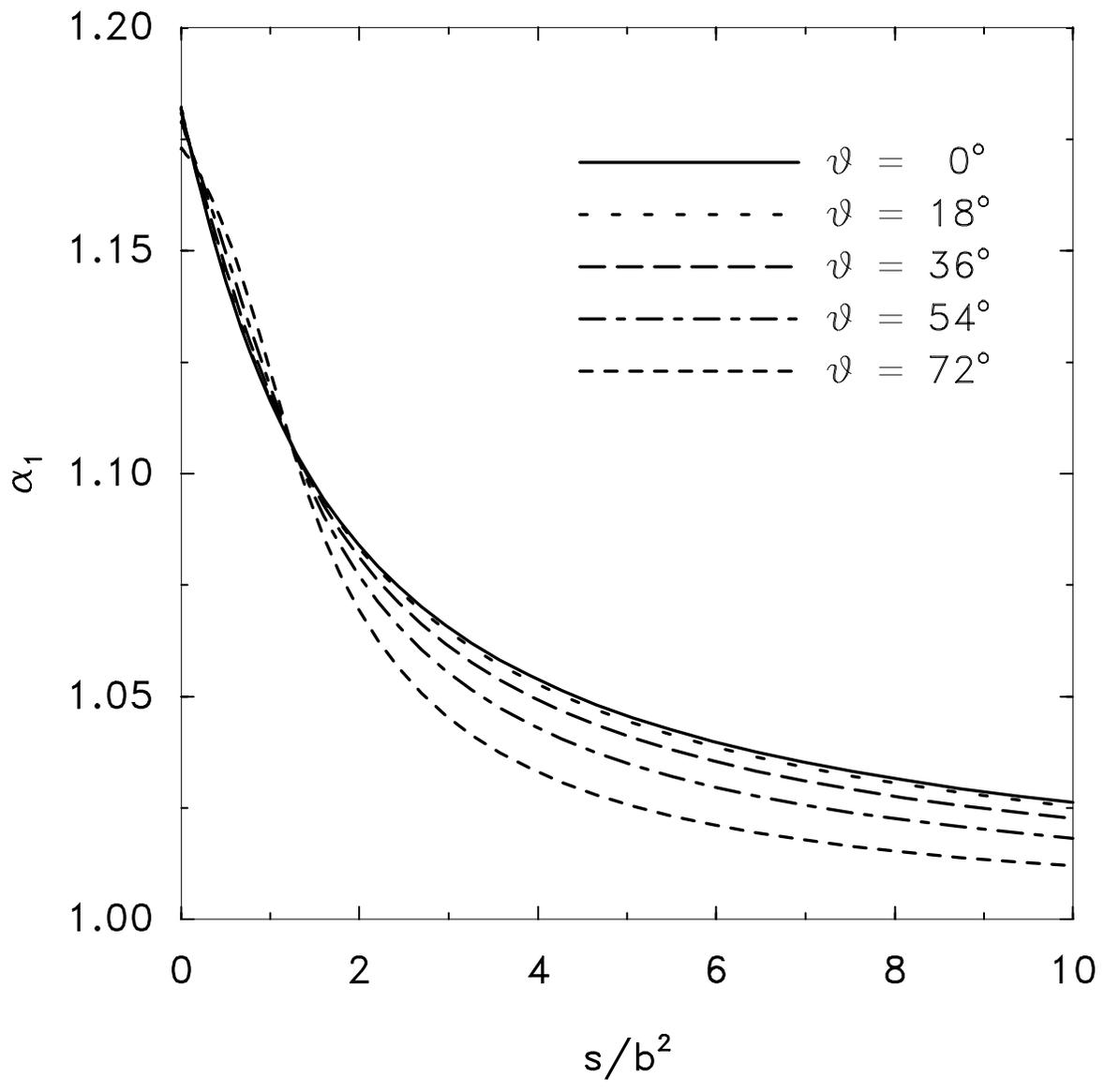
\end{document}